\documentclass[conference]{IEEEtran}
\IEEEoverridecommandlockouts
\usepackage{cite}
\usepackage{amsmath,amssymb,amsfonts}
\usepackage{algorithmic}
\usepackage{array,multirow}
\usepackage{graphicx}
\usepackage{textcomp}
\usepackage{xcolor}
\usepackage{url}
\usepackage{amsfonts}
\usepackage{subfigure}
\usepackage{booktabs}
\def\BibTeX{{\rm B\kern-.05em{\sc i\kern-.025em b}\kern-.08em
    T\kern-.1667em\lower.7ex\hbox{E}\kern-.125emX}}
\begin{document}

\title{SUNet: Swin Transformer UNet for Image Denoising\\
}


\author{\IEEEauthorblockN{Chi-Mao Fan and Tsung-Jung Liu}
\IEEEauthorblockA{Department of Electrical Engineering\\
National Chung Hsing University\\
Taichung 40227, Taiwan\\
Email: qaz5517359@gmail.com; tjliu@dragon.nchu.edu.tw}
\and
\IEEEauthorblockN{Kuan-Hsien Liu}
\IEEEauthorblockA{Department of Computer Science and Information Engineering\\
National Taichung University of Science and Technology\\
Taichung 40401, Taiwan\\
Email: khliu@nutc.edu.tw}}

\maketitle
\begin{abstract}
Image restoration is a challenging ill-posed problem which also has been a long-standing issue. In the past few years, the convolution neural networks (CNNs) almost dominated the computer vision and had achieved considerable success in different levels of vision tasks including image restoration. However, recently the Swin Transformer-based model also shows impressive performance, even surpasses the CNN-based methods to become the state-of-the-art on high-level vision tasks. In this paper, we proposed a restoration model called \emph{SUNet} which uses the Swin Transformer layer as our basic block and then is applied to UNet architecture for image denoising. The source code and pre-trained models are available at \url{https://github.com/FanChiMao/SUNet}.
\end{abstract}
\vspace{10pt}

\section{Introduction} \label{sec.1}
Image restoration is an important low-level image processing which could improve the performance in the high-level vision tasks, such as object detection, image segmentation and image classification. In the general restoration task, a corrupted image $Y$ could be represented as: 
\begin{equation}
Y=D(X) + n, \label{eq1}
\end{equation}
where $X$ is a clean image, $D(.)$ denotes the degradation function and $n$ means the additive noise. Some common restoration tasks are denoising, deblurring and deblocking. 

Traditional image restoration methods usually are based on algorithms, called prior-based or model-based methods, such as BM3D \cite{dabov2006image}, WNNM \cite{gu2014weighted} for denoising; deconvolution \cite{krahmer2006blind}, image prior \cite{shi2013single} for deblurring. Although most of convolution neural network (CNN)-based methods have achieved excellent performances \cite{ronneberger2015u,b26,b27,b28,b29,b31}, the naive convolution layer has several problems. First, the convolution kernel is content-independent with the images. Using the same convolution kernel to restore different image regions may not be the best solution \cite{liu2021swin,liang2021swinir}. Second, because the convolution kernel could be regarded as a small patch where the acquired features are local information, in other words, the global information will be lost when we do the long-range dependency modeling. Though in some papers, they proposed the methods to overcome the defects like adaptive convolution \cite{niklaus2017video,wu2021contrastive}, non-local convolution \cite{wang2018non} and global average pooling \cite{zhang2018image}, etc., they do not effectively solve the problems until the appearance of Swin Transformer.

Recently, \cite{liu2021swin} presented the new backbone based on transformer called Swin Transformer, and achieved the impressive performance on image classification. In addition, in more and more computer vision tasks including image segmentation \cite{wen2020identifying,chang2020locating,liu2021swin,cao2021swin}, object detection \cite{liu2021swin}, inpainting \cite{su2019image}, and super-resolution \cite{chen2019image,liang2021swinir}, using Swin Transformer as the backbone has surpassed the CNN-based methods to achieve the state-of-the-art. In this paper, we also consider Swin Transformer as our main backbone and integrate it into the UNet architecture called SUNet for image denoising.

Overall, the main contributions of this paper can be summarized as follows:

\begin{itemize}
\item We proposed a Swin Transformer network based on the image segmentation Swin-UNet model for image denoising. 
\end{itemize}
\begin{itemize}
\item We proposed a dual up-sample block architecture which comprises both subpixel and bilinear up-sample methods to prevent checkboard artifacts. The experiment results proved that it is better than the original up-sample from transpose convolution.
\end{itemize}
\begin{itemize}
\item To the best of our knowledge, our model is the first one to incorporate \emph{Swin Transformer} and UNet in denoising.
\end{itemize}
\begin{itemize}
\item We demonstrate the competitive results of our SUNet in two common datasets for image denoising.
\end{itemize}

\begin{figure*}[!htbp]
\centering
	\includegraphics[width=17.5cm]{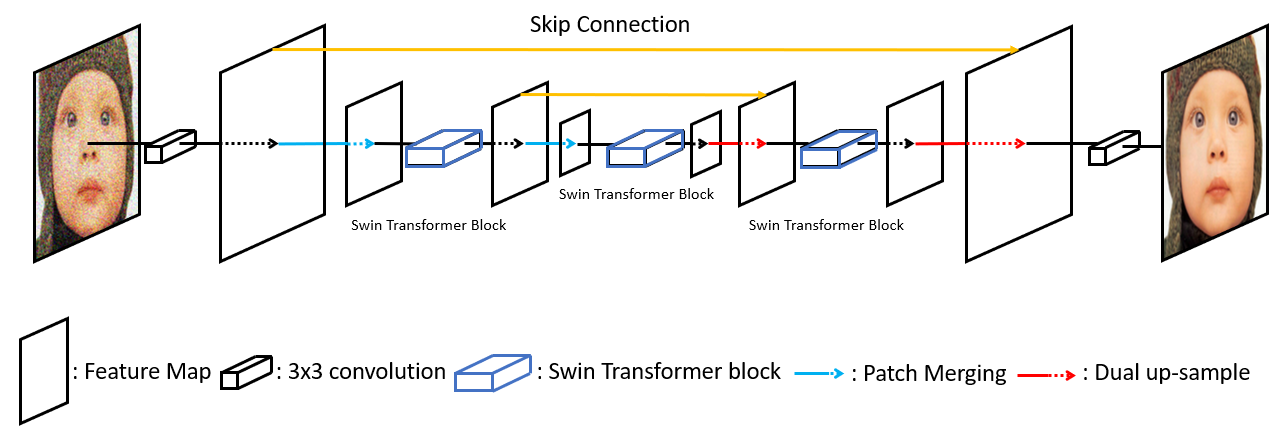}
	\caption{Proposed Swin Transformer UNet (SUNet) architecture. We first use $3 \times 3$ convolution to get the shallow feature. Then, they pass through the main feature extraction UNet. We use Swin Transformer Block as the basic extraction module to replace the naive convolution layer and acquire the high-level semantic information. For simplicity, the above figure only displays 2 layers of Swin Transformer Block, and the SUNet totally has \textbf{5 layers}. Finally, $3 \times 3$ convolution is used to reconstruct the restored image.}
	\label{SUNet}
\end{figure*}

\section{Related Work}
With the rapid development of hardware (e.g. GPU), the learning-based methods defeat the conventional model-based methods in both execution speed and performance. In this section, we first are going to introduce previous works about denoising. Then, we will describe the related works of UNet and Swin Transformer.
\subsection{Image Restoration} As aforementioned, traditional image restoration approaches are based on image priors or algorithms generally called model-based methods, such as self-similarity \cite{dabov2006image,buades2005non}, spare coding \cite{dong2011sparsity,xu2013unnatural} and total variation \cite{chan1998total}. The performance of these methods are acceptable on the ill-posed problem, but they have some shortcomings, such as time-consuming, computationally expensive, and difficult to restore complex image textures. Compared to conventional restoration methods, learning-based methods, especially convolution neural networks (CNNs) have become the mainstream in the computer vision field including image restoration because of the impressive performance. 

\subsection{UNet} Nowadays, UNet \cite{ronneberger2015u} is a well-known architecture in a lot of applications of image processing since it has hierarchical feature maps to gain the rich multi-scale contextual features. In addition, it uses the skip connection between encoders and decoders to enhance the reconstruction process of images. UNet is widely used in many computer vision tasks like segmentation, restoration \cite{park2019densely,b29}. Furthermore, it has various improved versions like Res-UNet \cite{res}, Dense-UNet \cite{dense}, Attention UNet \cite{att} and Non-local UNet \cite{nloc}. Due to the strong adaptive backbone, the UNet can be easily applied with different extractive blocks to enhance the performance. 

\subsection{Swin Transformer} \label{sec2.c}
Transformer \cite{b17} model is successful in the natural language processing (NLP) area and also has competitive performances with CNNs especially on image classification \cite{b18,b19}. However, the two main problems of directly using transformer to vision tasks are: 1) The difference of scale between images and sequences is large. The transformer has the defect of modeling the long sequence because it needs about square times of parameters of 1-dimension sequence. 2) Transformer is not good at solving the dense prediction tasks like instance segmentation which is a pixel-wise level task \cite{b20}. However, Swin Transformer \cite{liu2021swin} solves the above problems with shifted-window to decrease the parameters, and achieves the state-of-the-art performance in lots of pixel-wise vision tasks. 

\section{Proposed Method}

\subsection{SUNet}
The architecture of the proposed Swin Transformer UNet (SUNet) is based on the image segmentation model \cite{cao2021swin} and illustrated in Fig.~\ref{SUNet}. SUNet consists of three modules: 1) Shallow feature extraction; 2) UNet feature extraction; and 3) Reconstruction module.

\noindent\textbf{Shallow feature extraction module.} For a noisy input image $Y \in \mathbb{R}^{H \times W \times 3}$ where $H, W$ are the resolution of a corrupted image. We use single $3 \times 3$ convolution layer $M_{SFE}(.)$ to get the low-frequency information like color or texture of the input image. The shallow feature $F_{shallow} \in \mathbb{R}^{H \times W \times C} $ can be represented as:
\begin{equation}
F_{shallow} = M_{SFE}(Y), \label{eq2}
\end{equation}
where $C$ is the number of channels for shallow features, where we all set to 96 in the latter experiment section. 

\noindent\textbf{UNet feature extraction module.} Then, the shallow feature $F_{shallow}$ will be fed into the UNet feature extraction $M_{UFE}(.)$ to extract the high-level and multi-scale deep features $F_{deep} \in \mathbb{R}^{H \times W \times C}$:
\begin{equation}
F_{deep} = M_{UFE}(F_{shallow}), \label{eq3}
\end{equation}
where $M_{UFE}(.)$ is the UNet architecture with Swin Transformer Block, which contains 8 Swin Transformer Layers in single block to replace the convolutions. The Swin Transformer Block (STB) and Swin Transformer Layer (STL) will be illustrated with details in next subsection. 

\begin{figure}[htbp] 
	\hfill
	\subfigure[Swin Transformer Block (STB)]{
	
	\begin{minipage}[t]{1\linewidth}
	\centering
	\includegraphics[width=7cm]{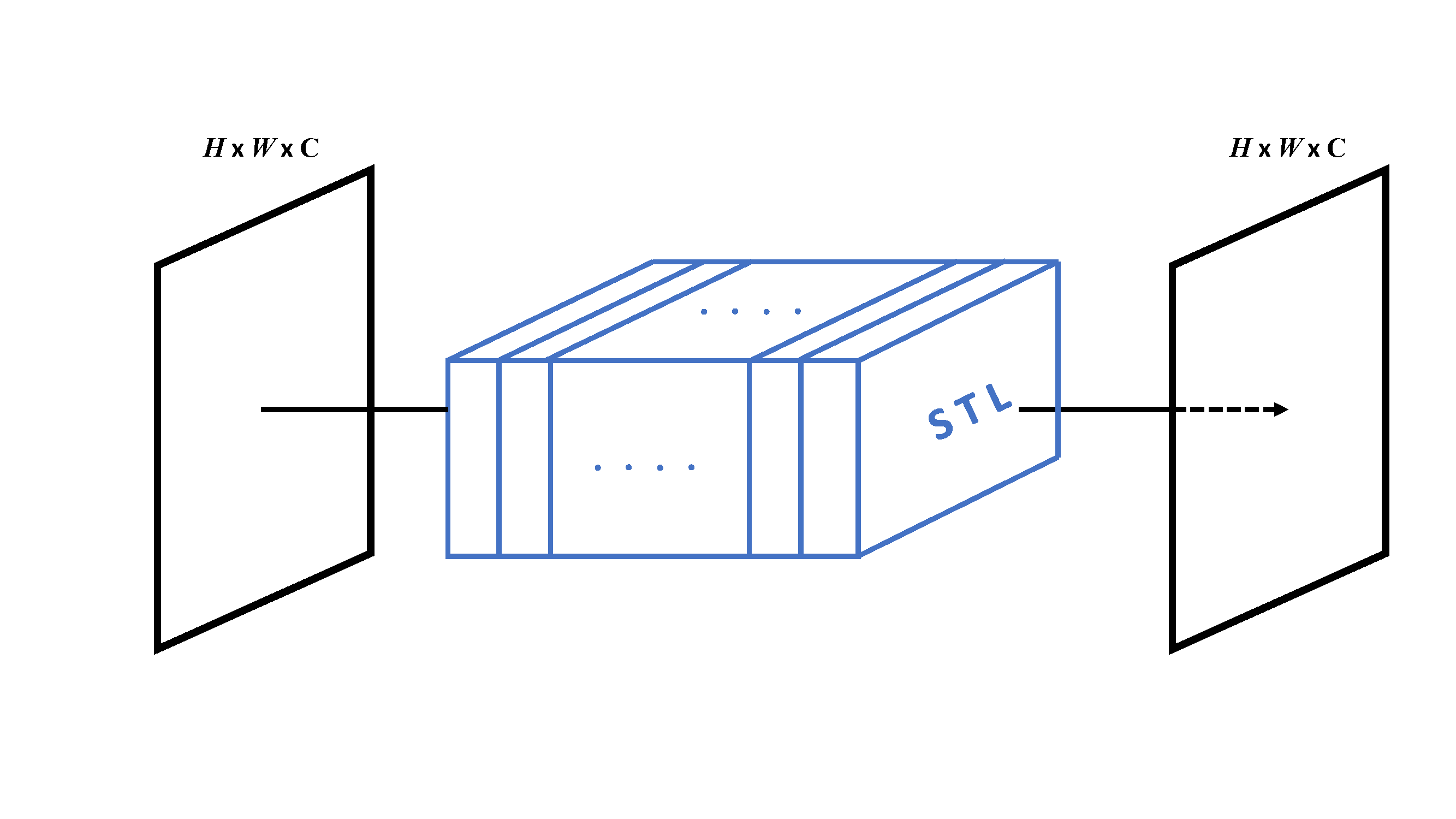}
	\label{STB}
	\end{minipage}%
	}%

	\subfigure[Swin Transformer Layer (STL)]{
	\begin{minipage}[t]{1\linewidth}
	\vspace*{-3.5mm}
	\centering
	\includegraphics[width=7cm]{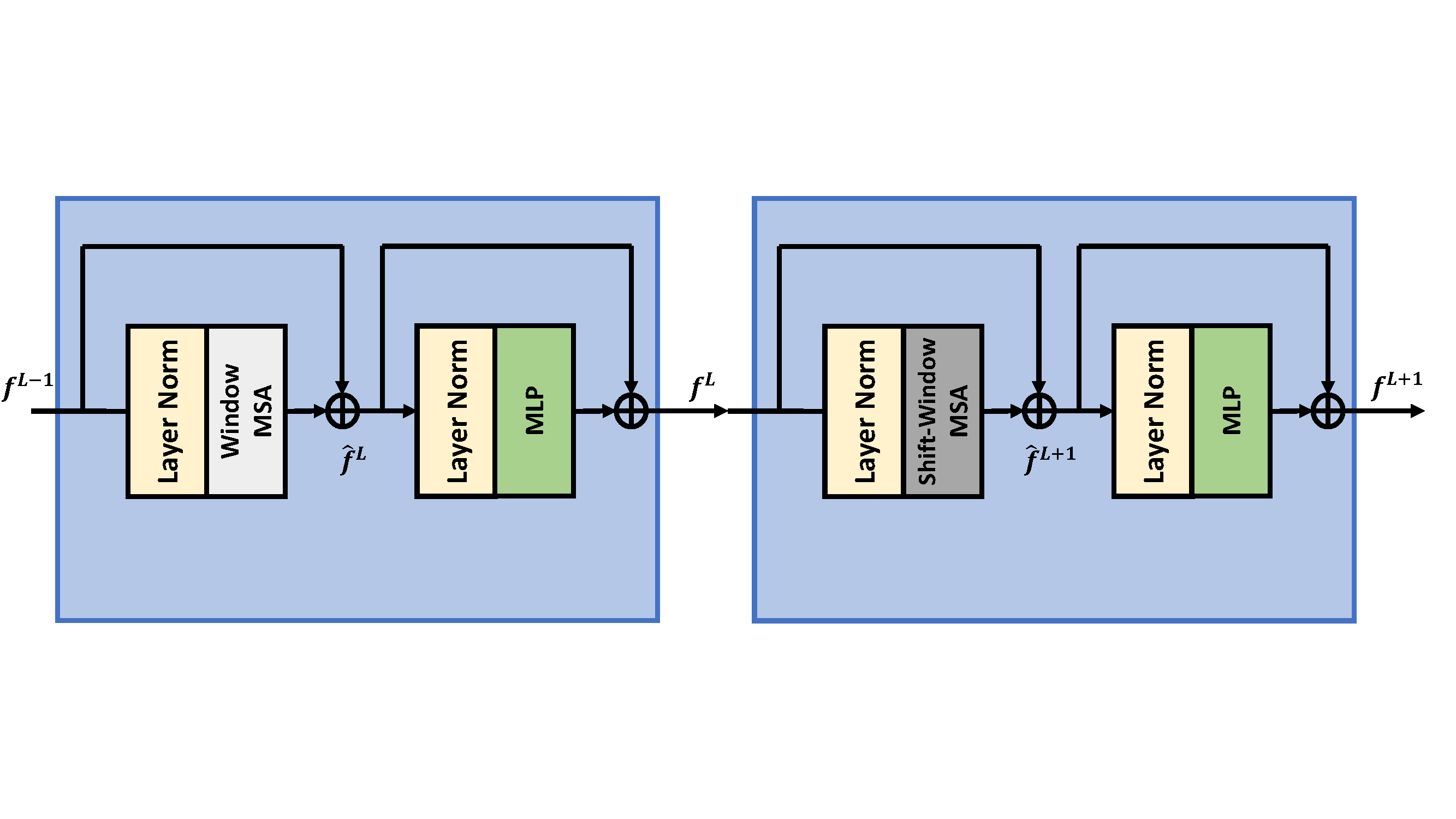}
	\label{STL}
	\end{minipage}%
	}%
\centering
\caption{(a) Swin Transformer Block (STB) which has 8 Swin Transformer Layers in our experiments. (b) Swin Transformer Layer (STL). Here, it has two STLs.}
\label{ST}
\end{figure}

\noindent\textbf{Reconstruction module.} Finally, we still use a $3 \times 3$ convolution $M_{R}(.)$ to generate the noise-free image $\hat{X} \in \mathbb{R}^{H \times W \times 3}$ from deep features $F_{deep}$ which is formulated as: 
\begin{equation}
\hat{X} = M_{R}(F_{deep}). \label{eq4}
\end{equation}
Note that $\hat{X}$ is obtained by taking the noisy image $Y$ as the input of SUNet and $X$ is the ground-truth and clean version of image of $Y$ in \eqref{eq1}.

\noindent\textbf{Loss function.} We optimize our SUNet end-to-end with the regular L1 pixel loss for image denoising:
\newcommand{\Lagr}{\mathcal{L}} 
\begin{equation}
 \Lagr_{denoise} =  ||\hat{X} - X||_1.  \label{eq5}
\end{equation}

\subsection{Swin Transformer Block} In UNet extraction module, we use STB to substitute the traditional convolution layer as shown in Fig.~\ref{ST}. STL \cite{liu2021swin} is based on the original Transformer layer \cite{b17} from NLP. The number of STL is always multiples of two, where one is for window multi-head self-attention (W-MSA), and the other is for shifted-window multi-head self-attention (SW-MSA). As mentioned in Section \ref{sec2.c}, there are some problems when directly using Transformer in CV tasks. Thus, they proposed the cyclic shift technique to decrease the computing time and keep the characteristics of convolution, including translation invariance, rotation invariance, and size invariance of the relationship between the receptive field and layers. Due to the page limits, we do not explain the principle of SW-MSA and how much computational complexity it could decrease in this paper. But we want to emphasize a key property of Swin Transformer (i.e., we could control the resolution ($H, W$) and channel number ($C$) of the output features as the same as the convolution operation). Taking Fig.~\ref{STL} for example, the whole process is represented as: 
\begin{equation}
\begin{aligned}
 \hat{f}^{L}& =  W{-}MSA(LN(f^{L-1})) + f^{L-1}, \\
         f^{L}& =  MLP(LN(\hat{f}^{L})) + \hat{f}^{L}, \\
 \hat{f}^{L+1}& =  SW{-}MSA(LN(f^{L})) + f^{L}, \\
 f^{L+1}& =  MLP(LN( \hat{f}^{L+1})) +  \hat{f}^{L+1}, \\ 
\end{aligned}
\end{equation}
where $LN(.)$ denotes as Layer Normalization, $MLP$ is multi-layer perceptron which has two fully connected layers with Gaussian Error Linear Unit (GELU) activation function. 

\subsection{Resizing module}
Since UNet has different scales of feature maps, the resizing modules (e.g., down-sample and up-sample) are necessary. In our SUNet, we use patch merging and proposed dual up-sample as the down-sample and up-sample module, respectively.

\noindent\textbf{Patch merging.} For down-sampling module, we follow \cite{liu2021swin,cao2021swin} to concatenate the input features of each group of $2 \times 2$ neighboring patches, and then use the linear layer to obtain the specified channel number of output features. We could also see this as the first step of doing the convolution operation, which is to unfold the input feature maps.

\begin{figure}[htbp]
\centering
	\includegraphics[width=8cm]{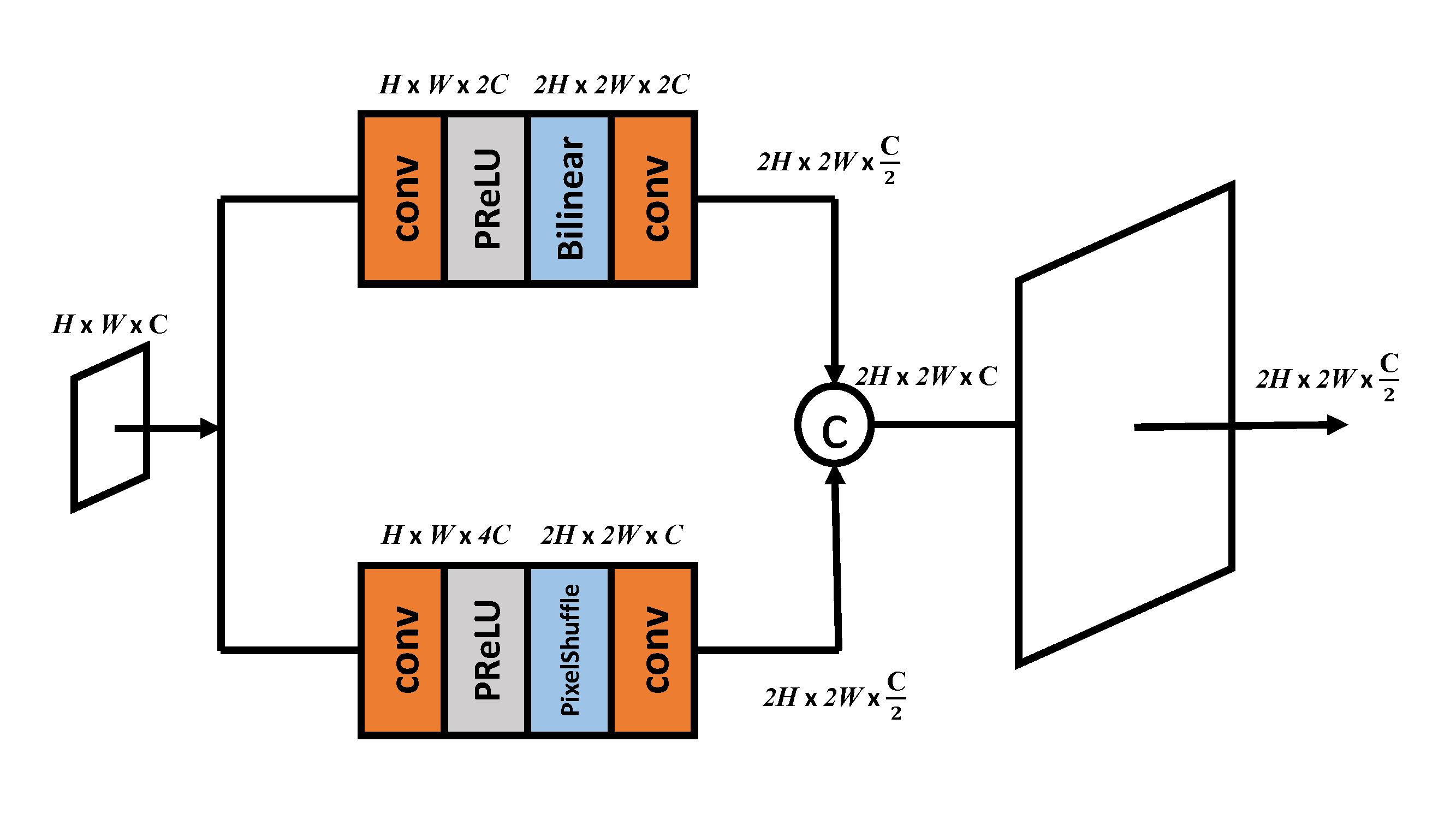}
	\caption{Proposed dual up-sample module with Bilinear and Sub-pixel up-sampling methods.}
	\label{dual_up}
\end{figure}

\noindent\textbf{Dual up-sample.} As for up-sample, the original Swin-UNet \cite{cao2021swin} uses patch expanding method which is equivalent to transpose convolution in the up-sampling module. However, the transpose convolution is easy to face the block effects. Here, we propose a new module called dual up-sample which comprises two existing up-sample methods (i.e., Bilinear and PixelShuffle \cite{b23}) to prevent checkerboard artifacts. The architecture of the proposed up-sampling module is shown in Fig.~\ref{dual_up}.

\begin{table*}[htbp] \footnotesize
\caption{Image Denoising results on CBSD68 dataset \cite{b24} and Kodak24 dataset \cite{b25}. Best and second best scores are \textbf{highlighted} and \underline{underline}, respectively. All of scores are the average values of the whole dataset. Last column of floating-point operations per second (FLOPs) is conducted on $256\times256$ color images.}
\label{denoise_table}
\begin{center}
\begin{tabular}{c  || cc | cc| cc || cc | cc| cc || cc}
\toprule[1.5 pt]
\multirow{3}{*}{Methods}&\multicolumn{6}{c||}{CBSD68 \cite{b24}}&\multicolumn{6}{c||}{Kodak24 \cite{b25}}&\multirow{3}{*}{Parms}&\multirow{3}{*}{FLOPs}\\
&\multicolumn{2}{c}{$\sigma = 10$}&\multicolumn{2}{c}{$\sigma = 30$}&\multicolumn{2}{c||}{$\sigma = 50$}&\multicolumn{2}{c}{$\sigma = 10$}&\multicolumn{2}{c}{$\sigma = 30$}&\multicolumn{2}{c||}{$\sigma = 50$}\\
&\multicolumn{1}{c}{PSNR}&\multicolumn{1}{c}{SSIM}&\multicolumn{1}{c}{PSNR}&\multicolumn{1}{c}{SSIM}&\multicolumn{1}{c}{PSNR}&\multicolumn{1}{c||}{SSIM}&\multicolumn{1}{c}{PSNR}&\multicolumn{1}{c}{SSIM}&\multicolumn{1}{c}{PSNR}&\multicolumn{1}{c}{SSIM}&\multicolumn{1}{c}{PSNR}&\multicolumn{1}{c||}{SSIM}\\
\midrule[1.5 pt]
Noisy         		&24.87&0.711&20.57&0.535&15.03&0.307&28.27&0.796&18.97&0.412&14.91&0.256&-&-\\
CBM3D \cite{CBM3D}
&35.89&0.951&29.71&0.843&27.36&0.763&33.32&0.943&27.75&0.773&25.60&0.686&-&-\\
UNet \cite{ronneberger2015u}
&35.39&0.948&29.74&0.849&27.35&0.771&35.89&0.939&30.55&0.845&28.11&0.774&17M&40G\\
DnCNN \cite{b26}
&36.12&0.951&30.32&0.861&27.92&0.788&36.58&0.945&31.28&0.858&28.94&0.792&558K&36G\\	
IrCNN \cite{b31}
&36.06&0.953&30.22&0.861&27.86&0.789&36.70&0.945&31.24&0.858&28.92&0.794&420K&27G\\	
FFDNet \cite{b27}
&\underline{36.14}&\underline{0.954}&\underline{30.31}&0.860&\underline{27.96}&0.788&36.80& 0.946 &31.39&0.860&29.10&0.795&854K&18G\\
DHDN \cite{b28}
&36.05&0.953&30.12&0.858&27.71&0.787&\textbf{37.30}& \underline{0.951} &\textbf{31.98}&\underline{0.874}&\textbf{29.72}&\underline{0.817}&168M&1019G\\
RDUNet \cite{b29}
&\textbf{36.48}&0.951&\textbf{30.72}&\textbf{0.872}&\textbf{28.38}&\textbf{0.807}&\underline{37.29}&0.901&\underline{31.97}&\underline{0.874}&\textbf{29.72}&\textbf{0.818}&166M&807G\\
\midrule[1.5 pt]
\textbf{SUNet (Ours)}
&35.94&\textbf{0.958}&30.28&\underline{0.870}&27.85&\underline{0.799}&36.79&\textbf{0.953}&31.82&\textbf{0.899}&\underline{29.54}&0.810&99M&30G\\
\bottomrule[1.5pt]
\end{tabular}
\end{center}
\end{table*}

\begin{figure*}[htbp]
	\centering
	\subfigure{
	\begin{minipage}{7cm}
	\includegraphics[width=6cm]{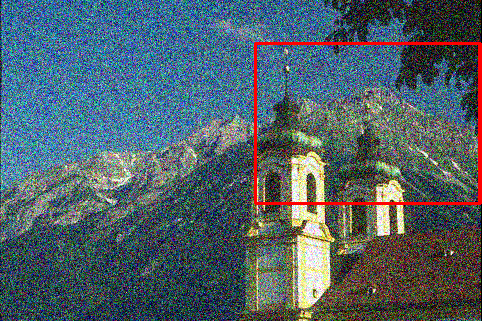}
	\centerline{\small{14.82/0.214}}
	\centerline{\small{Noisy Image}}
	\end{minipage}
	}
	\hspace*{-14mm}
	\subfigure{
		\begin{minipage}{10cm}
		\subfigure{
		\begin{minipage}{2cm}
		\includegraphics[width=1.9cm]{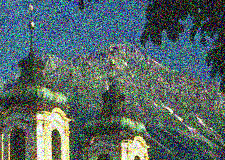}
		\centering
		\centerline{\small{14.80/0.299}}
		\centerline{\small{Noisy}}
		\end{minipage}%
		}%
		\subfigure{
		\begin{minipage}{2cm}
		\includegraphics[width=1.9cm]{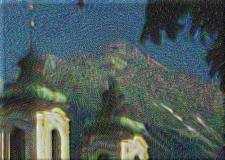}
		\centering
		\centerline{\small{20.57/0.475}}
		\centerline{\small{CBM3D \cite{CBM3D}}}
		\end{minipage}%
		}%
		\subfigure{
		\begin{minipage}{2cm}
		\includegraphics[width=1.9cm]{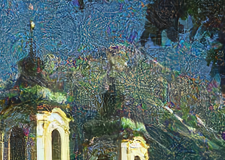}
		\centering
		\centerline{\small{21.23/0.435}}
		\centerline{\small{U-Net \cite{ronneberger2015u}}}
		\end{minipage}%
		}%
		\subfigure{
		\begin{minipage}{2cm}
		\includegraphics[width=1.9cm]{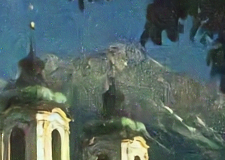}
		\centering
		\centerline{\small{24.97/0.694}}
		\centerline{\small{DnCNN \cite{b26}}}
		\end{minipage}%
		}%
		\subfigure{
		\begin{minipage}{2cm}
		\includegraphics[width=1.9cm]{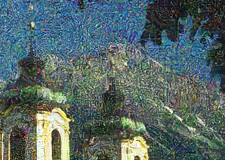}
		\centering
		\centerline{\small{20.58/0.393}}
		\centerline{\small{IrCNN \cite{b31}}}
		\end{minipage}%
		}%

		\subfigure{
		\begin{minipage}{2cm}
		\includegraphics[width=1.9cm]{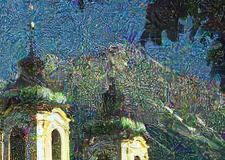}
		\centering
		\centerline{\small{20.72/0.417}}
		\centerline{\small{FFDNet \cite{b27}}}
		\end{minipage}%
		}%
		\subfigure{
		\begin{minipage}{2cm}
		\includegraphics[width=1.9cm]{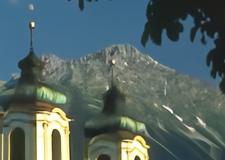}
		\centering
		\centerline{\small{\underline{27.48}/\textbf{0.812}}}
		\centerline{\small{DHDN \cite{b28}}}
		\end{minipage}%
		}%
		\subfigure{
		\begin{minipage}{2cm}
		\includegraphics[width=1.9cm]{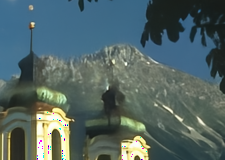}
		\centering
		\centerline{\small{26.24/0.812}}
		\centerline{\small{RDUNet \cite{b29}}}
		\end{minipage}%
		}%
		\subfigure{
		\begin{minipage}{2cm}
		\includegraphics[width=1.9cm]{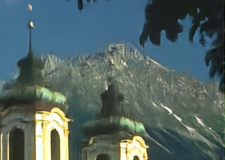}
		\centering
		\centerline{\small{\textbf{27.55}/\underline{0.811}}}
		\centerline{\small{SUNet (Ours)}}
		\end{minipage}%
		}%
		\subfigure{
		\begin{minipage}{2cm}
		\includegraphics[width=1.9cm]{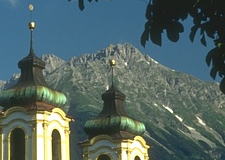}
		\centering
		\centerline{\small{PSNR/SSIM}}
		\centerline{\small{GT}}
		\end{minipage}%
		}%
		\end{minipage}	}%
	
\caption{Visual comparisons for image denoising on image '126007' from CBSD68\cite{b24} dataset corrupted by AWGN with $\sigma = 50$. The PSNR and SSIM values below the subfigures are calculated by patches.}
\label{denoising_images}
\end{figure*}

\section{Experiments}

\subsection{Experiment Setup}
\noindent\textbf{Implementation Details.} Our SUNet is an end-to-end trainable model without any pretrained networks and implemented by PyTorch 1.8.0 with single NVIDIA GTX 1080Ti GPU.

\noindent\textbf{Evaluation Metrics.} For the quantitative comparisons, we consider the Peak Signal-to-Noise Ratio (PSNR) and Structure Similarity (SSIM) Index metrics. Note that both PSNR and SSIM values are all the higher the better, and the unit of PSNR is decibel (dB). 

\subsection{Experiment Datasets}

\noindent\textbf{Training Set.} Using the same experimental setups of image denoising \cite{b28,b29}, we train our model on image super-resolution DIV2K \cite{b31} dataset which has 800 and 100 high-quality (the average resolution is about $1920\times1080$) images for training and testing, respectively. We randomly crop 100 patches with size of $256\times256$ for each training image and randomly add AWGN to the patches with noise level from $\sigma=5$ to $\sigma=50$ for 800 training images. As for validation, we directly use the testing set containing 100 images and add AWGN with three different noise levels $\sigma=10$, $\sigma=30$, and $\sigma=50$. 

\noindent\textbf{Testing Set.} For the evaluation, we choose CBSD68 dataset \cite{b24} which has 68 color images with the resolution of $768\times512$, and Kodak24 dataset \cite{b25} consisting of 24 images with the image size of $321\times481$.

\subsection{Image Denoising Performance}
We compare our SUNet with the prioir-based method (\emph{e.g.} CBM3D \cite{CBM3D}), CNN-based methods (\emph{e.g.} DnCNN \cite{b26}, IrCNN \cite{b31}, FFDNet \cite{b27}) and UNet-based methods (\emph{e.g.} UNet \cite{ronneberger2015u}, DHDN \cite{b28}, RDUNet \cite{b29}). Fig.~\ref{denoising_images} illustrates visual comparison \cite{liu2018study,liu2019modern} results for image denoising. In Table~\ref{denoise_table}, we conduct objective quality evaluation \cite{liu2012image,liu2015paraboost,liu2017no} of denoised image  and observe the following three things: 1) Our SUNet has competitive SSIM values because Swin-Transformer is based on the global information which makes the denoised images more perceptually faithful. 2) Compared to UNet-based methods (DHDN, RDUNet), the proposed SUNet has less parameters ($\downarrow 60\%$) and FLOPs ($\downarrow 3\%$) among the three models, and still keeps good scores on both PSNR and SSIM. 3) Compared with the CNN-based methods (DnCNN, IrCNN, FFDNet), we have the best PSNR and SSIM results among them along with almost the same FLOPs. Though the parameters of our model are the most (99M), it is caused by the self-attention operation which is not able to share the weights of kernels. However, it is more reasonable that features in different layers should use different kernel values as we discussed in Section \ref{sec.1}.

\section{Conclusion}
In this paper, we present the SUNet architecture which is based on the new backbone of Swin Transformer and achieve the competitive results on denoising. Furthermore, we propose the dual up-sample module to avoid the checkerboard artifacts. It is too early to say the Swin Transformer can replace the convolution. However, the potential of Swin Transformer still deserves to be expected in the future. Our future works are going to attempt more complex restoration tasks, such as real-world noise and real-world blur, while the model is still based on Swin-Transformer Layers.


\bibliographystyle{IEEEtran}
\bibliography{references}

\end{document}